\documentclass[traditabstract,letter]{aa}
\usepackage{amsmath}
\usepackage{txfonts}
\usepackage{longtable}
\usepackage{natbib}
\usepackage{graphicx}
\usepackage{ifthen}
\newcommand{\nix}[1]{}
\newcommand{\epsl}[01]{{Earth Plan. Sci. Lett.}}

\begin{document}
\title{Protostellar collapse: rotation and disk formation}
\author{Werner M. Tscharnuter \and Johannes Sch\"onke
\and Hans-Peter Gail \and Ekaterina L\"uttjohann}
\offprints{W.\,M. Tscharnuter,\\
\email{wmt@ita.uni-heidelberg.de}}
\institute{Zentrum f\"ur Astronomie (ZAH), Institut f\"ur
Theoretische Astrophysik (ITA), University of Heidelberg,
Albert-\"Uberle-Str.~2, 69120 Heidelberg, Germany}
\date{Received / Accepted }
%
\abstract{%
We present some important conclusions from recent calculations
pertaining to the collapse of rotating molecular cloud cores with
axial symmetry, corresponding to evolution of young stellar objects
through classes 0 and begin of class I.
Three main issues have been addressed: (1) The
typical timescale for building up a preplanetary disk -- once more
it turned out that it is of the order of one free-fall time which
is decisively shorter than the widely assumed timescale related to
the so-called ``inside-out collapse''; (2) Redistribution of
angular momentum and the accompanying dissipation of kinetic
(rotational) energy -- together these processes govern the
mechanical and thermal evolution of the protostellar core to a
large extent; (3) The origin of calcium-aluminium-rich inclusions
(CAIs) -- due to the specific pattern of the accretion
flow, material that has undergone substantial chemical and
mineralogical modifications in the hot ($\ga 900$\,K) interior of
the protostellar core may have a good chance to be advectively
transported outward into the cooler remote parts ($\ga 4$\,AU,
say) of the growing disk and to survive there until it is
incorporated into a meteoritic body.} 
%
%
\maketitle
%
\section{Introduction}\label{Sec:Intro}
Stars are thought to form from regions within  molecular clouds
that for some reason became gravitationally unstable and started
to collapse under the influence of their own gravitational
attraction \citep[see, e.g.,][]{Lar03,McK07}. It is clear that
angular momentum \citep[e.g.,][]{Bod95,Toh02} and magnetic fields
\citep[e.g.,][]{Pud08} play an important role during many stages
of protostellar collapse. Angular momentum is particular important
because it is responsible for the formation of accretion disks
that are the birthplaces of planetary systems. Therefore, the
study of rotating collapse is inevitable if one intends to
understand the formation of the Solar System and other planetary
systems.

The earliest such studies are a number of analytic approaches for  rotating
collapse without \citep{Ter84} and with \citep{All03} magnetic fields and
analytic studies of disk formation 
\citep[see][ and references therein]{Cas94}
and simple models for build-up and evolution of accretion disks
\citep[e.g.,][]{Lin90,Nak94}. These
suffer from that they introduce a number of approximations on the
nature of the collapse process that appear plausible but lack justification.

In the past two decades various attempts have been made, therefore, to follow
the 2-D and 3-D collapse of rotating protostellar fragments by means of
extended numerical calculations
\citep[e.g.,][to give a few examples]{Bos89,Bod90,Yor93,Sai08}
based on grid-methods, by a solution method based on a series
development into orthogonal polynomials \citep{Tsc87}, and by studies based
on the SPH-method \citep[e.g.,][]{Wal09}.
%
However, because of lack of spatial resolution almost all of these
studies followed only the first collapse phase up to the formation
of the first core. Neither the second collapse of the core to
stellar dimensions ($\la\!1\,\textrm{R}_{\sun}$) could be followed,
nor the formation of the inner part of the accretion disk on size
scales $\ll\!1\,\rm AU$. Only the model of \citet{Tsc87} and the
recent study of \citet{Sai08} achieved sufficient resolution to
follow the second collapse, but \citet{Sai08} neglects most of the
basic physics and considers only an over-simplified equation of
state.

Many of the investigations mainly focus on the individual stability
behaviour of the rapidly spinning, flattened, quasi-equi\-li\-brium
``cores'' that form in the innermost optically thick parts of
collapse flows. It has long been known that, for a rotating
self-gravitating equilibrium configuration, the ratio of rotational
over gravitational energy,
$\theta:=E_{\mathrm{rot}}/\left|E_{\mathrm{grav}}\right|$,%
\footnote{Usually
written $\beta:=T/\left|W\right|$, but we wish to avoid
confusion with $T$, the temperature, and $\beta$, the
scaling parameter of the turbulent viscosity.} %
is the appropriate parameter to indicate the onset of secular, for
$\theta\simeq0.14$, and dynamical, for $\theta\simeq0.27$,
instability. This is a standard result for \emph{homogeneous}
equilibria, but interestingly enough, in a series of papers
\citealp{Bod70,Bod73,Ost73} were able to show that these two
critical numbers still hold for more general rotating polytropes,
almost independently of the polytropic index, $n$.

Due to this remarkable finding extensive 3-D calculations have been
conducted in order to explore the fate of rapidly spinning ``naked''
protostellar cores by following the onset of the dynamical
instability and its development into the nonlinear regime. Much
effort has gone into clarifying the growth and saturation of the
self-gravitating non-axisymmetric modes (bars, spirals, mixed forms)
and their mutual interactions which drive a very effective
redistribution of the cores' angular momentum
\citep[e.\,g.,][]{Pic96,Tom98,Pic97,Ima00}.

In this letter we present some results of a new model calculation
that follows for the first time simultaneously the evolution of a
rotating Bonnor-Ebert-sphere-like initial state with
$\sim\!1\,\textrm{M}_{\sun}$ initial mass through axially-symmetric
first and second collapse to nearly stellar central densities and
the early build-up and evolution of the associated accretion disk.
The maximum resolution achieved amounts to about
$0.03\,\textrm{R}_{\sun}$ in the central parts. The adaptive grid is
chosen in a way that the innermost rigidly rotating homogeneous
sphere always contains a fraction of $10^{-6}$ of the total mass.
The simulation covers the evolutionary phases of young stellar
objects corresponding to class 0 and early class~I. It clearly
demonstrates the rapid co-formation of a compact stellar object and
a very extended accretion disk within a period of no more than
$\sim\!1.1$ free-fall times. The model includes all of the essential
physics of the problem: hydrodynamics, gravity, radiative transfer
(in Eddington approximation), and a realistic equation of state
including effects of sublimation of ice and dust on the opacity.

\begin{table}[t]
\centering \caption{Starting parameters.}\label{Tab:InitModel}
\begin{tabular}{lcr@{$\cdot$}ll@{}}
  \hline\hline
  Quantity                  & Symbol                                                       & \multicolumn{2}{c}{Value} & Dimension\rule[-1ex]{0pt}{3.5ex} \\
  \hline
  mass                      & $M$                                                          & \multicolumn{2}{c}{1.037} & M$_{\sun}$\rule{0pt}{2.5ex} \\
  angular\,momentum         & $L$                                                          & 2.682&10$^{53}$           & g\,cm$^2$\,s$^{-1}$ \\
  radius                    & $R$                                                          & 1.200&10$^{17}$           & cm \\
  spin angul.\,veloc.       & $\Omega$                                                     & 3.160&10$^{-14}$          & s$^{-1}$ \\
  centrifugal barrier       & $R_{\mathrm{cfb}}$                                           & 1.504&10$^{15}$           & cm \\
  mean density              & $\bar{\rho}$                                                 & 2.850&10$^{-19}$          & g\,cm$^{-3}$ \\
  mean free-fall time       & $\bar{t}_{\mathrm{ff}}=\sqrt{\frac{3\pi}{32G\bar{\rho}}}$    & 1.247&10$^{5}$            & yr \\
  central density           & $\rho_{\mathrm{c}}$                                          & 5.326&10$^{-18}$          & g\,cm$^{-3}$ \\
  central free-fall time    & $t_{\mathrm{ff,c}}=\sqrt{\frac{3\pi}{32G\rho_{\mathrm{c}}}}$ & 2.885&10$^{4}$            & yr \\
  temperature               & $T$                                                          & \multicolumn{2}{c}{10}    & K  \\
  ratio rot./grav.\,energy  & $\theta=\frac{E_{\mathrm{rot}}}{|E_{\mathrm{grav}}|}$        & 2.437&10$^{-3}$           & -- \\[1.5ex]
  \hline
\end{tabular}
\end{table}

By using a fully implicit numerical method (see
Section~\ref{Sec:Methoden}) we are now able to follow the collapse
of a slowly rotating, slightly Jeans-unstable molecular cloud
fragment with axial symmetry.

\section{Methods}\label{Sec:Methoden}
The most frequently applied numerical methods -- such as nested
grids by \citeauthor{Sai08} or SPH by \citeauthor{Wal09} -- have in
common that they are of explicit type (``forward'' time differences),
i.\,e., they are subjected to the Courant-Friedrichs-Lewy (CFL)
condition for the  timestep  so as to warrant numerical stability. In fact, the
CFL-timestep becomes prohibitively low with respect to the accretion
time, if the structure of the hot stellar core is to be sufficiently
well resolved. This is why explicit methods of any kind are genuinely
unsuited for attacking the overall problem of stellar formation in
a consistent way.

\nix{
\begin{table}[t]
\centering \caption{Zero-age parameters for the innermost
sphere.}\label{Tab:Zentr_1}
\begin{tabular}{lcr@{$\cdot$}ll}
  \hline\hline
  Quantity          & Symbol              & \multicolumn{2}{c}{Value} & Dimension\rule[-1ex]{0pt}{3.5ex} \\
  \hline
  radius            & $r_{\mathrm{c}}$    & 2.654&10$^{13}$           & cm \rule{0pt}{2.5ex}\\
  mean density      & $\rho_{\mathrm{c}}$ & 2.539&10$^{-14}$          & g\,cm$^{-3}$ \\
  revolution period & $\Pi_{\mathrm{c}}$  & 2.223&10$^{4}$            & yr  \\
  temperature       & $T_{\mathrm{c}}$    & \multicolumn{2}{c}{10.15} & K   \\
  gas pressure      & $P_{\mathrm{c}}$    & 9.348&10$^{-12}$          & bar \\[0.5ex]
  \hline
\end{tabular}
\end{table}
}

Unfortunately, appropriate implicit methods (``backward'' time
differences) have been worked out only for 1-D and 2-D collapse
problems with spherical and axial symmetry, respectively
\citep[see, e.\,g.,][]{WuTs03,Tsc87}. Although implicit numerical
schemes do not suffer from the CFL-condition, they demand, after
all, the solution of a huge system of non-linear equations which
is computationally expensive. Thus, implicit methods are useful
only if the physical processes to be considered exhibit a
hierarchy of timescales and the system as a whole evolves into a
quasi-stationary state. If the longest timescale exceeds the
smallest one by a large amount, the use of implicit methods is
mandatory, as is in our case, where the accretion time is several
orders of magnitude longer than the oscillation period of the
stellar core.

For a detailed description of the method which the implicit 2-D
hydrodynamical code %
is based on we refer to \citet{Tsc87}. For the equation of state,
dissociation of H$_2$ and ionisation of H and He is taken into
account; the opacities at low temperatures are dominated by dust
with and without ice mantles. In a spherical polar coordinate
system, we discretize the variables on a staggered radial grid
(256 gridpoints) and represent the dependence on the polar angle
by choosing a Legendre expansion (up to 27 coefficients) for each
primary variable. The discretized equations are written in
conservation form on a self-adaptive radial grid according to
\citet{Dor87}, shock fronts are smoothed out by artificial
(tensor-)\,viscosity.

\begin{figure}[t]
\centering
\includegraphics[width=\linewidth]{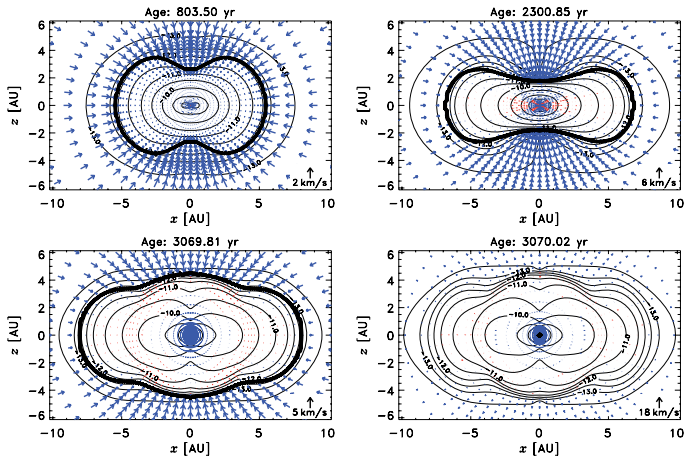}
\caption{Meridional cross sections. Displayed are equi-density
contours, the spacing being 0.5\,dex, and the velocity field. Red
``spots'' indicate expansion (cf.\ Fig.~\ref{fig:Kern-Aequ}a). The
closed heavy line marks the accretion shock, the arrow at the lower
right corner of the four panels represents the respective maximum
infall velocities. Upper row: snapshots slightly after B and before
C, respectively. Lower row: snapshots for D and E (cf.\
Fig.~\ref{fig:Kern}). At E (right panel) two shocks, an outer and an
inner one, bounding the growing ``pre-planetary'' disk and the
``stellar'' core, respectively, coexist (not displayed here, but
cf.\ Fig.~\ref{fig:Kern_2}).} \label{fig:Kern_1}
\end{figure}

Concerning the coefficient of turbulent viscosity,
$\nu_{\mathrm{tur}}$, we have adopted the so-called
$\beta$-viscosity prescription of \citet{Dus00}, an extension of
the well-known, but likewise heuristic,
$\alpha$-viscosity for application to self-gravitating disks. To
be specific, $\nu_{\mathrm{tur}}:=\beta r^2\Omega$, with $r$ being
the radial coordinate, $\Omega$ the (local) angular
velocity, and $\beta=10^{-4}\ldots 10^{-2}$ the
inverse of the critical Reynolds number indicating the onset of
turbulence. In our calculation we have chosen $\beta=10^{-2}$.

\section{Results and conclusions}\label{Sec:Resultate}

\begin{figure}[t]
\centering
\includegraphics[width=\linewidth]{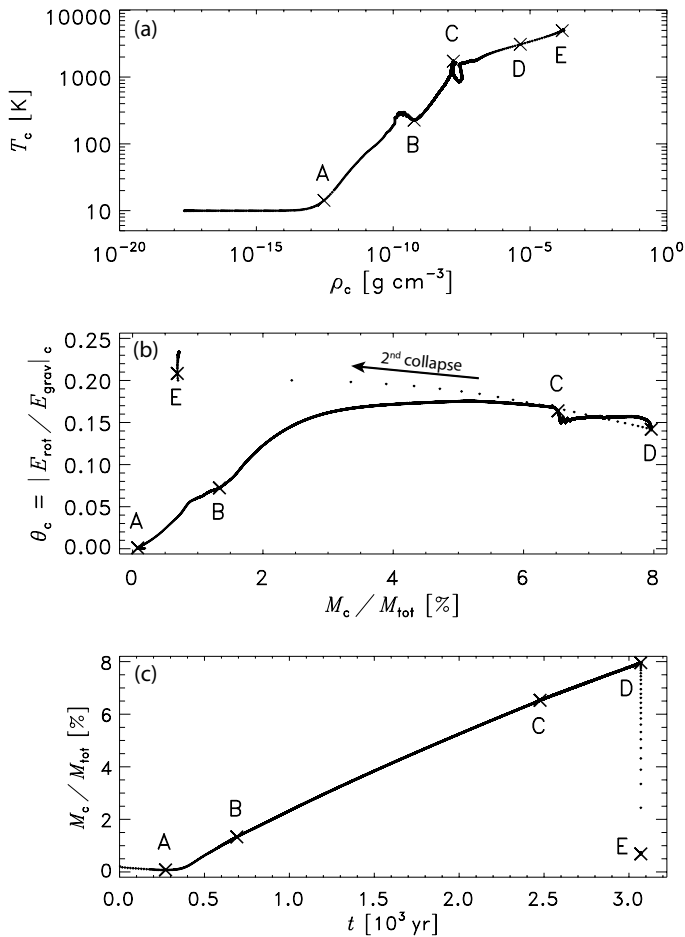}
\caption{Evolution of the core. (a): central density vs.\ central
temperature. (b) ratio of rotational over gravitational energy
vs.\ core mass. (c) core mass vs.\ time. For the meaning of labels
A through E see Section~\ref{Sub:Kernentwicklung}.}%
\label{fig:Kern}
\end{figure}

\subsection{Initial configuration and the first collapse phase}
\label{Sub:InitModel}
Table~\ref{Tab:InitModel} lists a set of appropriate starting
parameters leading to the collapse of a rotating protostellar
cloud fragment. The (optically thin) Bonnor-Ebert-sphere-like
initial configuration exhibits already a moderate density
concentration toward the centre and is assumed to rotate like a
rigid body. The revolution period, $2\pi/\Omega=6.3\cdot10^6$\,yr,
adopted in our calculation is large compared to the mean free-fall
time. Hence, the centrifugal barrier is situated at a distance
from the centre of only about 1\,\% of the cloud radius, and the
collapse flow develops almost perfectly with spherical symmetry
for most parts of the cloud.

After about one mean free-fall time the temperature rises in the
centre of the fragment and a quasi-hydrostatic core is going to form.
It is thus convenient to reset the clock to zero when the optical
depth of the collapsing fragment, counted from the outer edge to
the centre, exceeds 2/3 for the first time (e.\,g., \Citealp{WuTs03}).
This event may be attributed to the beginning of the protostellar
evolution proper. All ages subsequently given are relative to this instant.

Shortly after age zero an accretion shock forms, that marks the
natural boundary of a ``core'', a flattened quasi-hydrostatic
rotating structure that is pressure-supported parallel and, in
essence, centrifugally-supported perpendicular to the axis of
rotation.

As the ``core'' we shall refer to the innermost \emph{subsonic}
region of the collapsing fragment, regardless of whether or not the
accretion shock is present. During the main accretion phase the flow
becomes indeed highly supersonic and a strong shock is present at
any polar angle (cf. Fig.~\ref{fig:Kern_1}). However, the retarding
effect of the centrifugal forces significantly reduces the velocity
components perpendicular to the axis of rotation. Particularly close
to the equatorial plane, this effect will sooner or later in the
evolution weaken the local shock strength considerably or even make
the flow discontinuity disappear completely. This will happen within
only a few initial mean free-fall times. Then infall of matter from
the envelope is eventually going to cease, whereas the ``core''
gradually turns into a veritable protoplanetary accretion disk.

\begin{figure}[t]
\centering
\includegraphics[width=\linewidth]{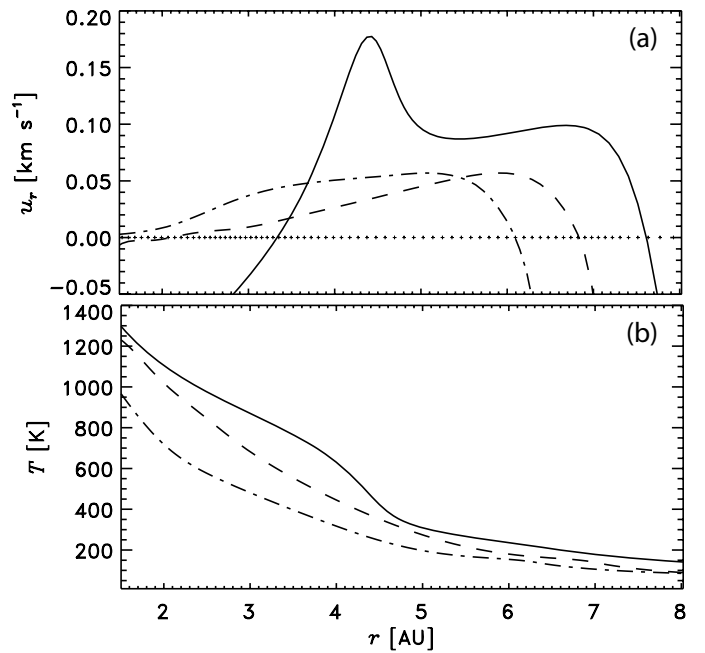}
\caption{Evolution of the core from its quasi-stationary stage to
the end of the second collapse. Spatial distribution of (a) the
radial velocity component, $u_r$, (b) the temperature, $T$, in the
equatorial plane for three instants of time indicated by dash-dotted
(2301\,yr), dashed (2948\,yr), and solid lines (3070\,yr),
respectively; crosses mark the gridpoints.} \label{fig:Kern-Aequ}
\end{figure}

The current simulations do not only cover the collapse proper, but
also the formation and the beginning of the ensuing accretion
phase of the disk-like core.

\subsection{Formation and growth of the disk-like core}
\label{Sub:Kernentwicklung}
Figure~\ref{fig:Kern} shows (a) the density-temperature-diagram
for the very centre of the collapsing cloud, (b) the stability
parameter $\theta_{\mathrm{c}}$ as a function of the relative core mass
$M_{\mathrm{c}}/M_{\mathrm{tot}}$, (c) the time dependence of the
relative core mass. The labels A through E mark characteristic
stages of the evolution:
\begin{itemize}
\item[A:] The core starts to grow in mass;
\item[B:] continuation of the accretion after thermal relaxation caused by
opacity effects (sublimation of ice mantels);
\item[C:] sublimation of the dust grains commences off-centre
near the axis of rotation as a ``hot polar cap'', leaving a
growing opacity gap and violent thermal relaxation effects;
\item[D:] start of the second collapse due to dissociation of H$_2$;
\item[E:] formation of the ``stellar'' core.
\end{itemize}

\begin{figure}[t]
\centering
\includegraphics[width=\linewidth]{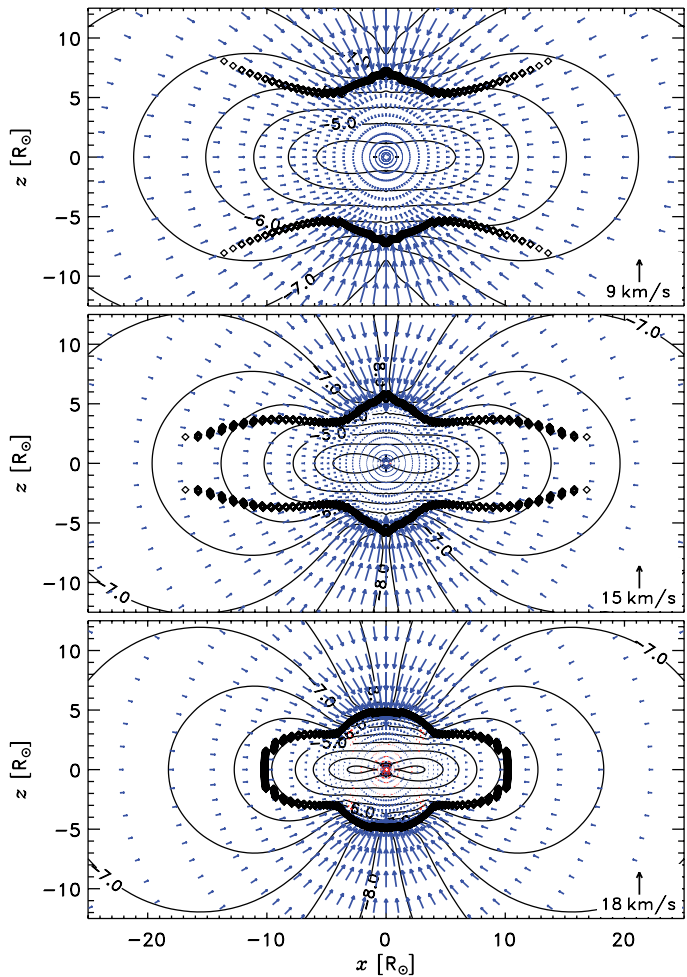}
\caption{Meridional cross sections (cf.\ Fig.~\ref{fig:Kern_1}).
Formation of the ``stellar'' core on a very short timescale. Plotted
are three snapshots of the final shock evolution at (from above)
$t_0-10$\,d, $t_0$, and $t_0+7$\,d, respectively, with
$t_0=3070$\,yr. Note that now the geometrical dimensions are only
several solar radii (R$_{\sun}$).}\label{fig:Kern_2}
\end{figure}

As the first important result, we have found a typical rise time,
$M_{\mathrm{c}}/\dot{M}_{\mathrm{c}}$, of the core's mass to be only
a small fraction (1--2\,\%) of the mean free-fall time, i.\,e., only
a few thousand years. Figure~\ref{fig:Kern}c shows an almost
constant accretion rate of $3\cdot10^{-5}$\,M$_{\sun}$\,yr$^{-1}$
lasting for about 2\,600\,yr, from slightly after ``A'' until ``D''
where the second collapse sets in. The dots between ``D'' and ``E''
represent the individual models tracing the dynamical transition
(the ``second collapse'') from the disk-forming first core to the
second ``stellar'' core.

Figure~\ref{fig:Kern_1} demonstrates the increasing geometrical
dimensions and the changing shape of the first core, which occurs
within a few thousand years; the bulge in the polar direction,
indicated by the route of the accretion shock in the vicinity of
the rotational axis, is the result of a rather intricate interplay
between redistribution of angular momentum, the accompanying
generation of entropy (viz.\ heat), and transport of energy. The
net effect are rising temperatures and lower rotation rates in the
inner parts of the core, which is necessary to trigger the second
collapse. As a matter of fact, test calculations have shown that
without a certain amount of angular momentum transport the core
would stay too cool and, hence, become prone to the onset of
destructive gravitational instabilities that trigger binary
formation.

A glimpse to Fig.~\ref{fig:Kern}b reveals that the first core cannot
escape from running into the regime of secular instability. The
ratio, $\theta=E_{\mathrm{rot}}/|E_{\mathrm{grav}}|$, quickly rises
above the critical number of 0.14, so that already a small amount of
dissipation will lead to symmetry breaking: The originally
axisymmetric core will take on a triaxial shape and presumably
evolve into a distinctive bar/spiral configuration exerting
gravitational torques which, in turn, results in an enhanced
redistribution of angular momentum. This finding may serve as
motivation for our rather efficient turbulent ($\beta$-)viscosity
with $\beta=0.01$.

\subsection{Formation of the ``stellar'' core}
\label{Sub:stellKern}
The second collapse is a result of the interplay between
thermodynamics, redistribution of angular momentum as a
dissipative process, and energy transport. Their combined effect
is illustrated by Fig.~\ref{fig:Kern_1}: particularly in the
central parts, angular momentum transport tends to defuse extrem
flattening, while the heat input by the accompanying dissipation
of rotational energy creates increasing pressure forces. As a
consequence, the pressure distribution is becoming more spherical,
and after some ``incubation'' period of slow contraction (between
``C'' and ``D'' in Fig.~\ref{fig:Kern}) dissociation of H$_{2}$
eventually leads to dynamical collapse (``D''--``E'').
Figure~\ref{fig:Kern}b shows that the stability parameter,
$\theta_{\mathrm{c}}$, approaches the critical number of 0.27
indicating dynamical instability. Interestingly enough, our
axisymmetric model suggests the formation of a self-gravitating
doughnut-like structure of the density distribution (cf.\
Fig.~\ref{fig:Kern_2}).

\subsection{An intermediate hot solar nebula?}
\label{Sub:stellHotNeb}
The model provides for the first time the initial temperature and
density structure of the accretion disk and the flow pattern in
the disk and its surroundings, including the accretion shocks, on
scales down to much less than 1\,AU. Figure~\ref{fig:Kern-Aequ}b
shows a rather hot inner disk region extending out to about 4\,AU.
This structure is not seen in earlier calculations because of lack
of resolution; it lasts for a couple of 1\,000\,yr. This inner
portion is hot enough for formation of materials like those found
as calcium-aluminium rich inclusions (CAIs) in meteorites; it
might be the ``hot solar nebula'' that cosmochemists always
advocated for.

From the onset of disk formation on there is both accretion and a
large-scale transport of matter (with velocities
10--50\,m\,s$^{-1}$) close to the disk's midplane from the hot part
outward to distances of several AU from the centre (cf.
Figs.~\ref{fig:Kern_1} and~\ref{fig:Kern-Aequ}a). Material from the
hot region may be mixed across the inner $\sim\!5$\,AU and part of
this material may survive till the onset of planetesimal formation,
since outflow close to the midplane continues to exist in later
phases, though with reduced velocity \citep[e.g.,][]{Kel04}. The
model therefore seems to offer an explanation how the short, maybe
as short as 20\,000\,yr \citep{Jac08}, CAI-forming period in the
Solar Nebula is related to the earliest evolutionary phase of the
accretion disk. If true, CAI formation follows immediately the
second collapse and is a valid indicator for the formation time of
the protoplanetary disk.

\begin{acknowledgements}
This work has been supported by the Forschergruppe 759, ``The
Formation of Planets: The Critical First Growth Phase'' of the
Deutsche Forschungsgemeinschaft (DFG).
\end{acknowledgements}

\end{document}